\documentclass{ws-procs9x6-cpt22}
\begin{document}

\newcommand{\refeq}[1]{(\ref{#1})}
\def\etal {{\it et al.}}

\title{Using ultra-high energy cosmic rays and air showers
to test Lorentz invariance
within modified Maxwell theory}

\author{Markus Risse}

\address{Department of Physics, University of Siegen,\\
57072 Siegen, Germany}

\begin{abstract}
Cosmic rays and air showers at ultra-high energy are unique tools to test the validity of Lorentz invariance.
A brief overview is given on such tests focusing on isotropic, non-birefringent Lorentz violation (LV) in the photon sector.
Based on the apparent absence of vacuum Cherenkov radiation and photon decay, the LV parameter $\kappa$ is bound to
$-0.6 \cdot 10^{-20} < \kappa < 6 \cdot 10^{-20}$ (98\% CL). We report an updated limit from cosmic-ray photon observations and
preliminary results on testing vacuum Cherenkov radiation in air showers.
\end{abstract}

\bodymatter

\section{Introduction}

As a pillar of physics, Lorentz invariance (LI) deserves being tested thoroughly.
The search for violation of Lorentz invariance (LV) is also motivated by efforts towards a fundamental theory, where LV effects are allowed and may appear in the low-energy theory.\cite{liberati09a}
Thus, theory needs experimental guidance. Data interpretation, in turn, needs a theoretical framework.

In the analyses summarized here, the framework of modified Maxwell theory is adopted (Sec.~\ref{sec:modmax}).
We make use of the highest-energy particles in the universe: ultra-high energy (UHE) cosmic rays (Sec.~\ref{sec:cr}) and the air showers they initiate (Sec.~\ref{sec:eas}).
By checking the presence of the non-standard processes of vacuum Cherenkov radiation and photon decay, LI is probed.
The apparent absence of these processes strongly constrains LV.


\section{Modified Maxwell theory}\label{sec:modmax}

The Lagrange density of standard QED is extended by adding a term which breaks Lorentz invariance while preserving CPT and gauge invariance\cite{modmax}:
\begin{equation}
\mathcal{L} = -\frac{1}{4}F^{\mu\nu}F_{\mu\nu} +
  \overline{\psi}\left[\gamma^\mu(i\partial_\mu-eA_\mu)-m\right]\psi
-\frac{1}{4}(k_F)_{\mu\nu\rho\sigma}F^{\mu\nu}F^{\rho\sigma}
\label{eq:lv_lagrangian}
\end{equation}
For a discussion, see Ref.~\refcite{lv2021} and references therein. We focus on the case of isotropic, nonbirefringent LV in the photon sector,
controlled by a single dimensionless parameter $\kappa \in (-1,1]$.
For $\kappa \neq 0$, certain processes forbidden in case of LI become allowed.
For $\kappa > 0$, vacuum Cherenkov radiation of charged fermions of mass $M$ occurs above an energy threshold
\begin{equation}
E^\text{th}_f(\kappa) = M\,\sqrt{\frac{1+\kappa}{2\kappa}} \simeq \frac{M}{\sqrt{2\kappa}}~.
\label{eq:particlethreshold}
\end{equation}
For $\kappa < 0$, photons decay into electron-positron pairs above the threshold
\begin{equation}
E^\text{th}_\gamma(\kappa) = 2\,m_e\,\sqrt{\frac{1-\kappa}{-2\kappa}} \simeq \frac{2\,m_e}{\sqrt{-2\kappa}}~,
\label{eq:photonthreshold}
\end{equation}
with the electron mass $m_e$.
Both processes turn out to be very efficient\cite{klinkhamer08c,diaz1516}: with radiation lengths $\ll 1$~m, the energy loss is quasi-instantaneous.


\section{Constraining Lorentz violation using cosmic rays}\label{sec:cr}


$\boldsymbol{\kappa > 0}$: A first constraint can be obtained by the mere existence of cosmic rays.
Assume $\kappa = 0.5 \cdot 10^{-10} \Rightarrow E^\text{th}_p \simeq 10^{14}$~eV for protons: 
then, no protons were expected to reach Earth above $E^\text{th}_p$, in contradiction to data.
Thus, $\kappa$ can be bounded. Assuming iron primaries for cosmic rays observed above $10^{20}$~eV
led to the constraint $\kappa < 6 \cdot 10^{-20}$ (98\% CL).\cite{klinkhamer08ab,klinkhamer08c}
\\ \\
$\boldsymbol{\kappa < 0}$: Similarly, the existence of cosmic-ray photons is used to constrain $\kappa < 0$.
Based on $\sim$30~TeV photons observed by atmospheric Cherenkov telescopes, a limit of
$\kappa > -9 \cdot 10^{-16}$ (98\% CL) was obtained.\cite{klinkhamer08c}
In view of the recent observation of PeV photons, this limit can be updated. Using a photon of (1.42$\pm$0.13) PeV energy\cite{lhaaso2021}
improves the limit to $\kappa > -4 \cdot 10^{-19}$ (98\% CL).
Cosmic-ray photons at even higher energies are searched for\cite{uhephoton1,uhephoton2} but have not yet been identified.
An observation of an EeV photon would improve the limit by about six orders of magnitude.


\section{Constraining Lorentz violation using air showers}\label{sec:eas}

Extensive air showers (EAS) initiated by cosmic rays offer an alternative approach to test LI:
in the cascading process, various UHE particles are expected to be produced as secondaries,
possibly with energies above $E^\text{th}_f$ or $E^\text{th}_\gamma$.
When the primary cosmic ray -- typically protons or nuclei up to iron -- interact with the air,
pions constitute a large fraction of the particles produced. The neutral pions usually decay quickly
into secondary photons, giving rise to the well-known electromagnetic cascade of photons, electrons and positrons
based on pair production and bremsstrahlung. Dominated by these electromagnetic particles, the number of particles
increases, reaches a maximum, and decreases again due to ionizational energy loss.

Air shower experiments such as the Pierre Auger Observatory\cite{auger} detect secondary particles reaching ground with a huge surface detector array
(3000~km$^2$). In addition, the flash of fluorescence light emitted by the air from the
passage of the particle cascade can be registered by appropriate telescopes.
This allows the observation of longitudinal shower profiles.
The energy of the primary particle is given by integrating the profile (modulo small corrections).
The atmospheric depth of shower maximum, $X_{\textrm{max}}$, contains information about the primary type:
due to the smaller cross-section,
primary proton showers reach the maximum about 100~g/cm$^2$ deeper in the
atmosphere compared to primary iron of same total energy.
Related, shower-to-shower fluctuations $\sigma(X_{\textrm{max}})$
are larger for primary protons. 
Both quantities, $X_{\textrm{max}}$ and $\sigma(X_{\textrm{max}})$, can be used to study possible LV effects.
\\ \\
%
%
%
\begin{figure}[t]
\begin{center}
\includegraphics[width=3.3in]{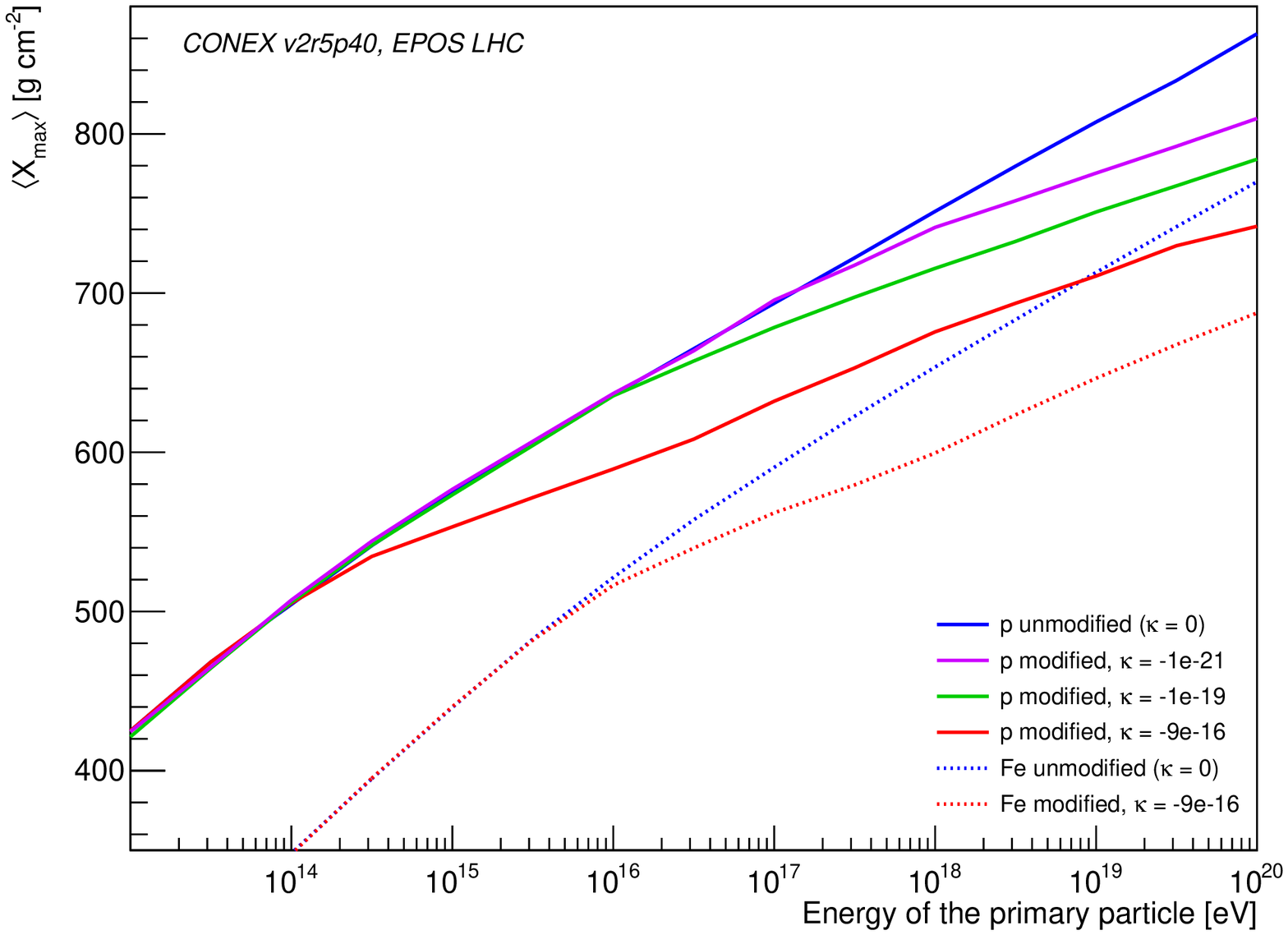}
\end{center}
\caption{Simulated $\left<X_{\textrm{max}}\right>$ vs.\ energy for $\kappa \le 0$ for primary proton and iron.\cite{lv2017}}
\label{fig1}
\end{figure}
$\boldsymbol{\kappa < 0}$: While awaiting the observation of primary UHE photons, 
in EAS secondary photons are expected with energies up to about 10\%
of the primary proton energy. This permits some access to photons well above PeV energies.
In case of quasi-instantaneous photon decay, showers become shorter\cite{lv2017}:
as shown in Fig.~\ref{fig1}, the average depth of maximum $\left<X_{\textrm{max}}\right>$ is reduced by $\sim$100~g/cm$^2$
at $10^{19}$~eV for $\kappa = -9 \cdot 10^{-16}$.
If the observed $\left<X_{\textrm{max}}\right>$ exceeds the one from LV simulations for any primary composition,
the corresponding $\kappa$ value is excluded.
%
%
This has been performed for $\left<X_{\textrm{max}}\right>$ only\cite{lv2017} and 
combining\cite{lv2021} $\left<X_{\textrm{max}}\right>$ and $\sigma(X_{\textrm{max}})$.
The combination provides additional restrictions on the primary composition and led to
a limit of $\kappa > -0.6 \cdot 10^{-20}$ (98\% CL). The constraints are summarized in Tab.~\ref{tbl1}.
\begin{table}[t]
\tbl{Summary of constraints on $\kappa$ (limits at 98\% CL).}
{\begin{tabular}{@{}l|rc@{}}
    & Using cosmic rays & Using air showers  \\
\colrule
	$\kappa > 0$: vacuum Cherenkov  & $< 6 \cdot 10^{-20}$ & (work in progress)  \\
\colrule
$\kappa < 0$: photon decay  & $> -40 \cdot 10^{-20}$ & $> -0.6\cdot 10^{-20}$ \\
\end{tabular}
}
\label{tbl1}
\end{table}
\\ \\
%
%
%
$\boldsymbol{\kappa > 0}$: Checking Tab.~\ref{tbl1}, the question arises whether also a limit on $\kappa > 0$ can be placed using EAS.
Secondary electrons and positrons with energies
up to about 5\% of the primary proton energy are expected. In case of instantaneous vacuum Cherenkov radiation, again shorter showers
could occur compared to the standard case of bremsstrahlung.
Preliminary results\cite{lv2022} after implementing this LV effect in shower simulations are displayed in Fig.~\ref{fig2}.
Indeed, again a reduction of $\left<X_{\textrm{max}}\right>$ can be noticed. The effect appears to be smaller
than the one for $\kappa < 0$ for the same $|\kappa|$. Still, a limit can be placed that is
based on a different method (showers instead of cosmic rays) and different particle species (electrons/positrons instead of protons/nuclei).
\begin{figure}[b]
\begin{center}
\includegraphics[width=3.3in]{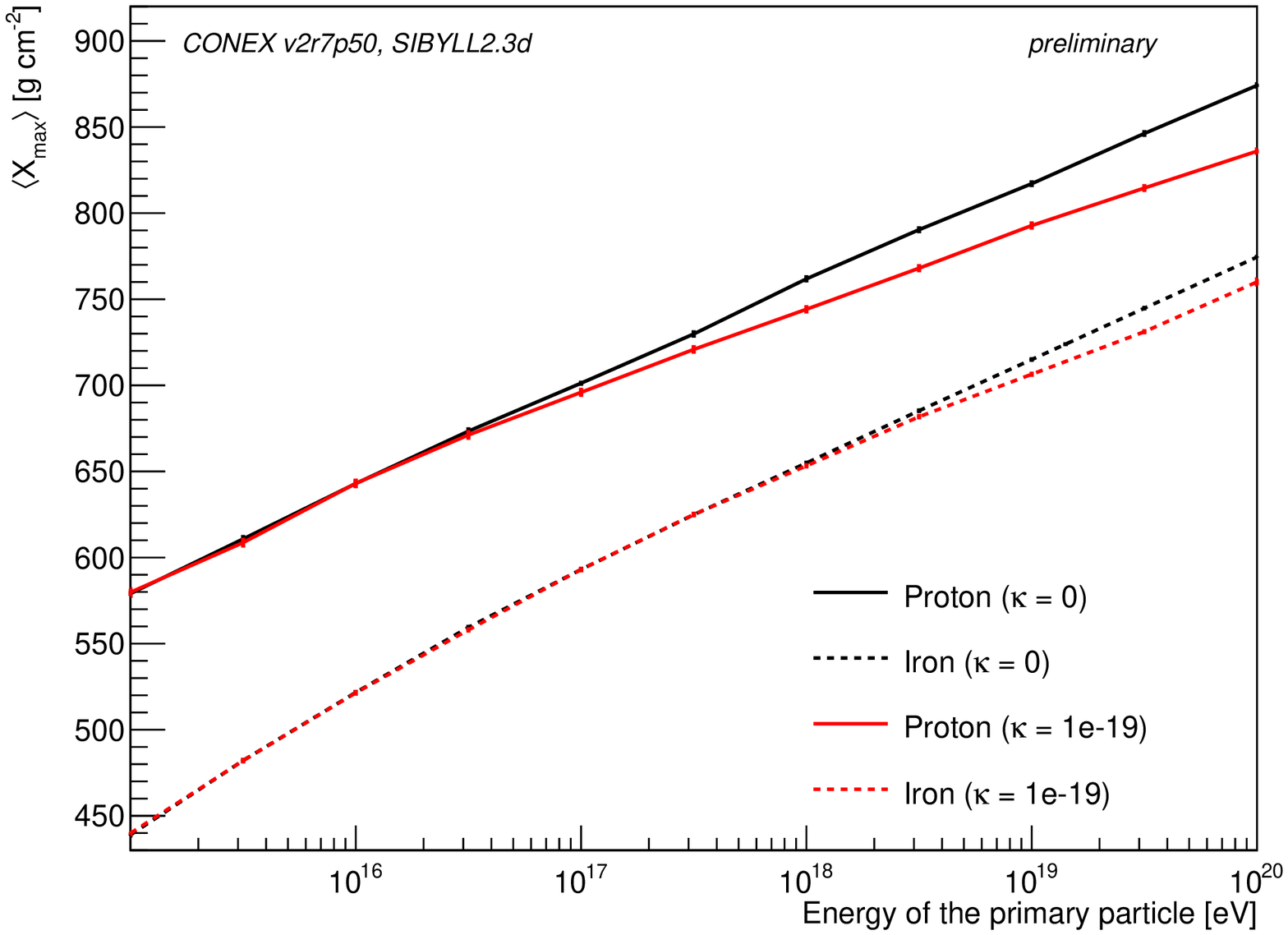}
\end{center}
\caption{Simulated $\left<X_{\textrm{max}}\right>$ vs.\ energy (prelim.) for $\kappa \ge 0$ for primary proton and iron.\cite{lv2022}}
\label{fig2}
\end{figure}
It should also be noted that for a given $\kappa$, only primaries below a certain
threshold $E/M$ arrive at Earth at all. This restriction on composition will strengthen the LV constraints.


\section{Conclusion and outlook}

UHE cosmic rays and their air showers are well suited to test LI.
The apparent absence of the non-standard processes of vacuum Cherenkov radiation and photon decay puts strong constraints on LV.
Currently, LV constraints on $\kappa > 0$ are derived using air showers.\cite{lv2022}
Further improved tests are possible in case of observing UHE photons and for further data (higher energy, smaller uncertainties, restrictions on composition).

With limits on $|\kappa|$ at the $10^{-20}$ level, the tested LI apparently holds to a high extent.
In view of theoretical approaches allowing a-priori for LV at much larger levels, the questions arises whether mechanisms
should be considered to ``protect LI'' (e.g., Ref.~\refcite{bjorken}).
Still, even if getting used to constraining LV, experimentalists should remain open-minded for LV showing up.

\section*{Acknowledgments}
It is a pleasure to thank the organizers for a stimulating workshop.
The work summarized here was performed in collaboration with F.R.\ Klinkhamer,
J.S.\ D\'iaz, M. Niechciol and F. Duenkel: thank you!
Support by the German Research Foundation (DFG) is gratefully acknowledged.

\end{document}